\documentclass[12pt]{article} 
\title{Forcing Free Fields}  
\author{{\it Richard Shurtleff~}\thanks{affiliation and mailing 
address: Department of Applied Mathematics and Sciences, 
Wentworth Institute of Technology, 550 Huntington Avenue, 
Boston, MA, USA, ZIP 02115, telephone number: (617) 989-4338, e-mail address: shurtleffr@wit.edu}} 
\begin{document} 

\maketitle

\begin{abstract} 
The momentum of a free massive particle, invariant under translation, thereby realizes a trivial representation of the translation group. By allowing nontrivial reps of translations, momentum changes with translation, a recipe for force. Here the procedure is applied to the conventional construction of a free quantum field using spacetime symmetries, yielding a more general field with the free field as a special case. It is shown that a particle described by the quantum field follows the classical trajectories of a massive charged particle in electromagnetic and gravitational fields. 



\end{abstract}

\section{Introduction}

For free particles, the symmetries of flat spacetime determine essentially unique quantum fields.\cite{W} Since not all particles are free, the derivation of free fields from spacetime symmetries may have hidden assumptions. By exploring alternatives to the implicit assumptions that lead to free fields, one might find fields that respond to forces. In this paper two implicit assumptions in the derivation of free fields are generalized and made explicit. Both assumptions involve translations.

Translation is a spacetime symmetry whose generators, the momentum operators, commute. Thus, by basic quantum theory, single particle states can be expressed in terms of momentum eigenstates whose eigenvalues are the `momentum'  $p.$ 

For free particles a translation of spacetime acts on a single particle state of momentum $p$ to produce a state with the same momentum $p.$ As a transformation applied to the momentum, the translation is trivial: $p$ goes into $p.$

There exist nontrivial reps of translations in which four-vectors, like a particle's momentum $p,$ change upon translation. The change of a particle's four-momentum as it moves summarizes the effects of force. Therefore it may be possible to obtain a quantum field that responds to forces by assuming a nontrivial rep for the effects of translations on a particle's momentum $p.$ 

However, in order to undergo nontrivial changes due to translations, the momentum must be accompanied by another multicomponent object, most generally a second rank tensor. The additional object is initially arbitrary and brings with it initially arbitrary forces that change momentum by translation. 

The implicit assumption that the translation rep is trivial is upgraded to an explicit assumption. Assumption 1, in Sec. \ref{Phi} allows a general translation rep for the momentum.

A second assumption is needed for quantum fields to feel the force of gravity. We note that the translation of spacetime by a displacement $b$ simply adds $b$ to the spacetime coordinates of every event in spacetime. This is a spacetime symmetry because the translation preserves coordinate differences:   
\begin{equation} \label{Deltax}
\Delta x = x_{2} - x_{1}\quad {\mathrm{and}} \quad \Delta x = (x_{2}+b) - (x_{1}+b) \, .\end{equation}
Any displacement leaves all coordinate differences unchanged. This is not true for rotations and boosts. With rotations and boosts, coordinate differences do change; it is their spacetime scalar products that are unchanged.

The creation and annihilation operators that underly the construction of quantum fields realize representations of the Poincar\'{e} group, which includes translations. Rotations and boosts must be the same for the operators as for spacetime in order to preserve spacetime scalar products.

As is well known, a translation along a displacement $\epsilon$ for annihilation and creation operators is represented by a phase factor $\exp{(\pm i p\cdot \epsilon)}.$ Since any displacement leaves all coordinate differences of spacetime unchanged, one may question whether the displacement $b$ of spacetime need be the same as the displacement $\epsilon.$ The $\epsilon$ in the phase factor becomes a free variable. 

The implicit assumption that the displacement $b$ of spacetime and the displacement $\epsilon$ for annihilation and creation operators are equal is replaced in this article by an explicit assumption, Assumption 2 in Sec. \ref{Qfields}, that allows the two displacements to differ, $\epsilon \neq b.$

Assumption 2 has an effect on the derivation of quantum fields presented in Sec. \ref{Qfields}. There one finds that the Invariant Coefficient Hypothesis determines the displacement $\epsilon.$ The displacement $\epsilon$ for the operators depends on the Poincar\'{e} transformation $(\Lambda , b)$ of spacetime and on the location $x$ in spacetime. The result is  $\epsilon$ = $\Lambda \epsilon_{0} - \Lambda M x + M^{\prime} x^{\prime},$ where $\epsilon_{0}$ is constant, $M$ is an arbitrary second rank tensor field, and $M^{\prime}$ is the field evaluated at $x^{\prime}$ = $\Lambda x + b.$ Thus a Poincar\'{e} transformation of spacetime $(\Lambda,b)$ is accompanied by an annihilation and creation operator representation of the Poincar\'{e} transformation $(\Lambda,\epsilon).$ 

One main result of the derivation in Sec. \ref{Qfields}, is that the expansion over plane waves $\exp{(\pm i p\cdot x)}$ familiar for free fields is generalized to waves $\exp{(\pm i p\cdot Mx)},$ which depend on the arbitrary tensor $M.$ 

The translation of momentum from event to event in Sec. \ref{Phi} and the field derived in Sec. \ref{Qfields} at a single event combine in Sec. \ref{ClassLimit} to determine the expected classical trajectories. Paths that make the phase extreme are found. These are the expected classical trajectories.\cite{Feynman}

For a given $M,$ the collection of expected classical trajectories form a system of `extreme coordinates' $X.$ Since $M$ is arbitrary it can be the identity and for $M$ = 1, we get a special extreme coordinate system $Y.$ It is shown in Sec.~\ref{ClassLimit} that $M$ determines the transformation matrix $\partial{Y}/\partial{X}$ from one set of extreme coordinates to the other, $M$ = $\partial{Y}/\partial{X}.$ 

Extreme coordinates undergo general transformations $M \rightarrow$ $N$ because the arbitrary tensor field $M$ can be replaced with some other tensor field $N.$ And the arbitrary property of $M$ follows from the freedom to choose distinct displacements $b$ and $\epsilon$ for spacetime and the operators. So general covariance in the curved spacetime of extreme coordinates traces back to the fact that any displacement of flat spacetime $x$ leaves all coordinate differences invariant, Eq.~(\ref{Deltax}).

In Sec. \ref{ClassLimit}, it is shown that the classical trajectories are those of charged massive particles in background electromagnetic and gravitational fields. That these are not arbitrary forces but electromagnetism and gravity follow from the requirement that the magnitude of the momentum is the constant particle mass. These electromagnetic and gravitational fields must be identified as background fields because the work presented here does not include any way of determining electromagnetic fields from charges and currents or any way to obtain the gravitational field from mass and velocity distributions. 

The deductions here result from applying Poincar\'{e} invariance to the invariant coefficient method of constructing quantum fields as sums of annihilation and creation operators. So other principles upon which quantum fields are interpreted such as equal time commutation rules that yield spin-statistics results are not considered here. They may be treated elsewhere.

\pagebreak
\section{Poincar\'{e} Transformations of Momentum } \label{Phi}

This section contains a summary of the Poincar\'{e} group of transformations of four-vectors such as momentum. The rule relating translations and equivalent momentum eigenstates is stated as Assumption 1 near the end of the section. 

Let $\Psi_{p,\sigma}$ be a single particle eigenstate of momentum,
\begin{equation} \label{eigen}
     P^{\mu}_{(\Psi)} \Psi_{p,\sigma} = p^{\mu} \Psi_{p,\sigma} \, ,
\end{equation}
where $P_{(\Psi)}$ is the momentum operator for the representation of the Poincar\'{e} group realized by the spacetime symmetries of the states $\Psi_{p,\sigma}$ and $\sigma$ is the $z$-component of spin. 

The generators $\{J_{(\Psi)},P_{(\Psi)}\} $ of the representation realized by the symmetries of the states $\Psi_{p,\sigma}$ have no direct relationship with the generators $\{J_{(\Phi)},P_{(\Phi)}\} $ of the representation that realizes the symmetries of four-vectors, defined below.   They are different representations of the same group. This article involves many representations of the Poincar\'{e} group and the various representations have different momentum and angular momentum operators.

Recall that the way a multicomponent object transforms with rotations and boosts determines its spin composition. Four-vectors have spin $(1/2,1/2).$ However, the spacetime symmetries of the Poincar\'{e} group include also translations and that implies that a spin $(A,B)$ field must be combined with fields of allowed spins $(A\pm 1/2,B\pm 1/2)$ in order to have nontrivial translation reps.\cite{rs0,L} For a four-vector $p$ with spin $(A,B)$ =  $(1/2,1/2),$ $p$ must be combined with objects of spin $(0,0)$ or $(0,1)$ or $(1,0)$ or $(1,1)$ in order to transform nontrivially upon translation.

But these spins are the spin composition of second rank tensors. Thus the four components of a four-vector $p$ can be combined with the components of a second rank tensor $T$ to create a multicomponent object $\Phi$ that can have nontrivial translations. We construct
\begin{equation} \label{pT1}
 \Phi_{m}  = \pmatrix{p^{\alpha} \cr T^{\gamma \delta}_{(\Phi)} } \, ,
\end{equation}
where the index $m$ runs from 1 to 20 since the momentum $p^{\mu}$ has 4 components and the tensor $T^{\gamma \delta}$ has 16 components. The indices $\alpha,\gamma,\delta \in$ $\{1,2,3,4\}$ indicate Minkowski coordinates, with $x^{4}$ = $t$ as time, so $p^{4}$ is energy. One can transcribe the 16 double index values $\gamma \delta$ = $\{11,12,\ldots, 44\}$ to the 16 single index values $m$ = $\{5,6,\ldots,20\}.$ 

Any Poincar\'{e} transformation $(\Lambda,\delta x)$ can be written as a homogeneous Lorentz transformation $\Lambda$ followed by a translation through a displacement $\delta x.$ The 20-component object $\Phi$ changes by a square matrix representation that acts on the 20 components of $\Phi_{l},$ 
\begin{equation} \label{transf2}
 \Phi_{l}^{\prime}  = \sum_{\bar{l}} D^{(\Phi)}_{l \bar{l}}(\Lambda,\delta x)  \Phi_{\bar{l}} \, ,
\end{equation}
where $D^{(\Phi)}(\Lambda,\delta x)$ is the matrix representing the spacetime transformation $(\Lambda,\delta x).$ 

Let $J_{(\Phi)}^{\mu \nu}$ be the $20 \times 20$ matrices for the angular momentum and boost generators of rotations and boosts and let $P_{(\Phi)}^{\mu}$ be the four $20 \times 20$ momentum matrices. Then the transformation matrices are determined by the displacement $\delta x_{\mu}$ and the antisymmetric parameters $ \omega_{\mu \nu}$ of the Lorentz transformation $\Lambda,$
\begin{equation} \label{transf1}
 D^{(\Phi)}(\Lambda,\delta x)  = \exp{(- i \delta x_{\sigma} P_{(\Phi)}^{\sigma})}\exp{( i \omega_{\mu \nu} J_{(\Phi)}^{\mu \nu}/2)}  \, ,
\end{equation}
where the exponent of a matrix $A$ is defined to be the infinite series, $\exp{A}$ = $\sum A^{n}/n!.$

The generators can be put in block matrix form with angular momentum and boost generators $J$ in the form
\begin{equation} \label{gen1}
 J_{(\Phi)}^{\rho \sigma}  = \pmatrix{(J^{\rho \sigma}_{11})^{\mu}_{\nu} && 0 \cr 0 && (J^{\rho \sigma}_{22})^{\alpha \beta}_{\gamma \delta} } \, ,
\end{equation}
where the 11-block generators are the usual ones for 4-vector transformations of $x^{\mu},$ 
\begin{equation} \label{J11} 
 (J^{\rho \sigma}_{11})^{\mu}_{\nu}  = i \left(\eta^{\sigma \mu} \delta^{\rho}_{\nu} - \eta^{\rho \mu} \delta^{\sigma}_{\nu} \right) \, ,
\end{equation} 
where $\eta^{\alpha \beta}$ is the metric diag$\{+1,+1,+1,-1\}$ and $\delta^{\rho}_{\nu}$ is the delta function, one for $\rho$ = $\nu$ and zero otherwise. The 22-block generators are usual for second rank tensors, e.g. $x^{\mu}_{1}x^{\nu}_{2},$
$$ 
 (J^{\rho \sigma}_{22})^{\gamma \delta}_{\epsilon \xi}  = -i \left(\eta^{\rho \gamma} \delta^{\sigma}_{\epsilon} \delta^{\delta}_{\xi}- \eta^{\sigma \gamma} \delta^{\rho}_{\epsilon} \delta^{\delta}_{\xi} + \eta^{\rho \delta} \delta^{\sigma}_{\xi} \delta^{\gamma}_{\epsilon}- \eta^{\sigma \delta} \delta^{\rho}_{\xi} \delta^{\gamma}_{\epsilon} \right) \, .
$$ 
These matrices can be found in the literature.\cite{WeinbergV}

There are two incompatible sets of momentum matrices, those with the 12-block nonzero that we use and those with the 21-block nonzero that we do not use. A convenient formula for the 12-block components is,
$$ 
 (P^{\sigma}_{12})^{\alpha}_{\gamma \delta}  = \hspace{14cm} $$ $$ i [ \left( C_1 + C_2 - C_3 \right)\delta^{\sigma}_{\gamma} \delta^{\alpha}_{\delta}  + 
 \left( C_1 + C_2 + C_3 \right)\delta^{\sigma}_{\delta} \delta^{\alpha}_{\gamma}  -  2 C_1 \eta^{\sigma \alpha} \eta_{\gamma \delta}
  + C_4  \eta^{\sigma \rho} \eta^{\alpha \kappa} \epsilon_{\rho \kappa \gamma \delta}]  \, ,
$$ 
where the constants $C_{i}$ have dimensions of an inverse distance. The expressions for the generators have appeared previously in an article that showed how free particles can be made to respond to electromagnetism and gravitation by upgrading the translation rep for momentum in classical mechanics.\cite{RS} 

Thus, the momentum matrices $P_{(\Phi)}^{\sigma}$ used to translate the quantity $\Phi$ have the form
\begin{equation} \label{gen4}
 P_{(\Phi)}^{\sigma}  = \pmatrix{0 && (P^{\sigma}_{12})^{\alpha}_{\gamma \delta} \cr 0 && 0 } \, .
\end{equation}
One can show that the generators $J_{(\Phi)}^{\mu \nu}$ and $P_{(\Phi)}^{\mu}$ satisfy the Poincar\'{e} algebra. 

Since our interest in this rep involves the effect of translation on a four-vector, set the Lorentz transformation to the identity, $\omega_{\mu \nu}$ = 0. By (\ref{transf1}) and (\ref{gen4}), we find that
%\begin{equation} \label{Dp1}
$$D^{(\Phi)}(1,\delta x) \Phi  = \exp{(- i \delta x_{\sigma} P_{(\Phi)}^{\sigma})} \Phi = (1 - i \delta x_{\sigma} P_{(\Phi)}^{\sigma}) \Phi  = \pmatrix{p^{\alpha} + \delta x_{\sigma} \bar{T}^{\alpha \sigma }_{(p)}\cr T^{\mu \nu}_{(\Phi)} } \, , $$
%\end{equation}
where the exponential simplifies by (\ref{gen4}) because the products of $P_{(\Phi)}$s vanish, $P_{(\Phi)}^{\mu} P_{(\Phi)}^{\nu}$ = 0, and where  $\bar{T}^{\alpha \sigma }_{(p)}$ is given by
$$\bar{T}^{\alpha \sigma }_{(p)} \equiv -i  (P^{\sigma}_{12})^{\alpha}_{\beta \gamma} T^{\beta \gamma}_{(\Phi)}
  \, . $$
The bar in $\bar{T}_{(p)}$ is placed there because the notation $T_{(p)}$ is reserved for another version of the tensor.

Thus a translation along a displacement $\delta x$ changes the four-vector momentum by the rule
\begin{equation} \label{Dp2}
{p^{\prime}}^{\alpha}  = p^{\alpha}+ \eta_{\sigma \mu} \bar{T}^{\alpha \sigma }_{(p)}  \delta x^{\mu}  \, .
\end{equation}
For a displacement $\delta x$ that is sufficiently small, the translated momentum $p^{\prime}$ is uniquely determined. For a large displacement one may construct many sequences of sufficiently small displacements. In general, the translated momentum $p^{\prime}$ depends on the particular sequence selected. This is something like path dependence, but since each sequence of displacements translates all of spacetime, there are infinitely many similar paths for each sequence.

{\it{Assumption 1: For a sufficiently small displacement of spacetime, the equivalence of the eigenstate of the translated momentum $p^{\prime}$ to the eigenstate of the initial momentum $p$ realizes a representation of the translation. }}

Since the momentum operators $P^{\mu}_{(\Psi)}$ for particle states in (\ref{eigen}) are invariant under translations, the set of eigenstates is the same before and after a translation. However the eigenvalues can change due to a translation. Indeed, Assumption 1 determines the post-translation eigenstate with eigenvalue $p^{\prime}$ in the (invariant) set of eigenstates that is equivalent to the pre-translation eigenstate with eigenvalue $p.$ 

In this section the displacements, not individual events are central. A translation in this section acts to move all events to other events, all displaced equally. In the next section a translation at a given event passively changes the coordinates of that event; the event is unchanged, only its coordinates change. We combine the two types of translations in Sec. \ref{ClassLimit}. 

\pagebreak
\section{Quantum Fields; Event-By-Event } \label{Qfields} 

In this section, we work at a single given event. There are Poincar\'{e} transformations, but these only provide the given event with new coordinate labels. The change in momenta in (\ref{Dp2}) is reserved for active translations that map events onto other events. The momenta in this section change with rotations and boosts to maintain scalar products with other four-vectors, but the momenta do not change with translations.  

At each event in spacetime a value for a quantum field may be constructed as a sum of annihilation and creation operators with invariant coefficients, the so-called Invariant Coefficient Hypothesis. The procedure here adapts a well-known derivation\cite{W} of free quantum fields of arbitrary spin.

Let the given event have coordinates $x$ in an initial reference frame. The quantum field $\psi_{l}(x)$ is defined to be a linear combination of the annihilation and creation operators $a_{\sigma}({\overrightarrow{p}})$ and $a^{\dagger}_{\sigma}({\overrightarrow{p}})$ which remove or add a single particle state $\Psi_{p,\sigma}$ with momentum $p$ and spin component $\sigma$ to any multiparticle state, see Eq. (\ref{eigen}). 

The annihilation and creation operators are defined over the appropriate Wigner class of momenta, $p^2$ = $-m^2$ with $p^{t} \geq m > 0$ and $m$ is the particle mass. The energy $p^{t}$ = $p^{4}$ can be determined from the spatial three-vector $\overrightarrow{p} ,$ $p^{t}$ = $(m^2+\overrightarrow{p}^2)^{1/2},$ so the operators depend on the three spatial components of $\overrightarrow{p}.$ 

It should be clear that working at a given event $x$ in no way confines the single particle states $\Psi_{p,\sigma}$ added or removed by the operators. 
Indeed the operators $a_{\sigma}({\overrightarrow{p}})$ and $a^{\dagger}_{\sigma}({\overrightarrow{p}})$ remove or add the same eigenstate at whatever event that one constructs $\psi_{l}(x);$ the operators are the same at all events in spacetime. 

However, the states $\Psi_{p,\sigma}$ which $a_{\sigma}({\overrightarrow{p}})$ and $a^{\dagger}_{\sigma}({\overrightarrow{p}})$ remove or add must be defined over suitably vast expanses of spacetime, in keeping with the Heisenberg uncertainty principle. A hidden assumption here is that the spacetime over which the single particle states $\Psi_{p,\sigma}$ are spread out is the same as the spacetime containing the given event $x.$

The quantum field $\psi_{l}(x)$ can be separated into linear combinations of annihilation and creation operators, $\psi_{l}(x)$ = $\kappa \psi^{+}_{l}(x)$ + $\lambda \psi^{-}_{l}(x),$ where the annihilation field $\psi^{+}$ and the creation field $\psi^{-}$ are given by 
%\begin{equation} \label{psi+}
$$\psi^{+}_{l}(x) = \sum_{\sigma} \int d^3 p \enspace u_{l\sigma}(x,{\overrightarrow{p}}) a_{\sigma}({\overrightarrow{p}})  \, ,$$
%\end{equation}
\begin{equation} \label{psi+}
\psi^{-}_{l}(x) = \sum_{\sigma} \int d^3 p \enspace v_{l\sigma}(x,{\overrightarrow{p}}) a^{\dagger}_{\sigma}({\overrightarrow{p}})  \, .
\end{equation}
Since the fields are linear combinations of operators, quantum fields are also operators. We reserve the term `operator' for the annihilation and creation operators. Thus operators are defined over the domain of allowed momenta and a field depends on the coordinates of the given event. The coefficients, the  $u$s and $v$s, depend on both.

The coefficients $u$ and $v$ can be determined from the ways the fields and operators transform to preserve spacetime symmetries: ({\it{i}}) the operators transform under Poincar\'{e} transformations with a unitary representation (rep), ({\it{ii}}) the coefficients are required to be invariant under Poincar\'{e} transformations and ({\it{iii}}) the quantum field transforms by a nonunitary rep. Remarkably, the unitary transformations of the operators $a$ and $a^{\dagger}$ can produce a non-unitary transformation of the fields $\psi^{\pm}_{l}$. The three transformation regimens constrain the coefficients so much that the coefficients are essentially determined.

It is assumed in other such calculations that the displacement applied to operators is identical to the displacement applied to the coordinates of the given spacetime event $x.$ However, all coordinate differences are unchanged by any translation. Applying different translations to operators than event coordinates is therefore allowed because either way spacetime coordinate differences remain invariant. 

The same does not hold for rotations and boosts. We must apply the same rotations and boosts to operators as those applied to event coordinates to preserve scalar products.

To replace the implicit assumption that operators and spacetime coordinates be translated by equal displacements we have the following explicit assumption:
 
{\it{Assumption 2. Suppose the spacetime coordinates $x$ of the given event transform with a combination $\Lambda$ of rotations and boosts followed by a translation through a displacement $b,$ in symbols: $x \rightarrow$ $\Lambda x + b.$ Assume that the annihilation and creation operators transform with the same Lorentz transformation $\Lambda,$ but the displacement $\epsilon$ for the operators is a suitably differentiable, coordinate-dependent function $\epsilon$ of the spacetime transformation $(\Lambda,b).$ }}

We write the function as $\epsilon (\Lambda,x,b).$ The displacement $\epsilon$ could have been chosen to be arbitrary, since any displacement preserves all coordinate differences. Assumption 2 allows the displacement $\epsilon$ to be different at different coordinates $x$ for the given event and for different transformations $(\Lambda,b).$ More general choices may be considered elsewhere.  

%\pagebreak

The annihilation and creation operators transform by a unitary representation of the Poincar\'{e} transformation $(\Lambda,\epsilon).$ We use a notation similar to that in Ref. \cite{W},
%\begin{equation} \label{Da+}
$$U(\Lambda,b) a_{\sigma}({\overrightarrow{p}}) {U}^{-1}(\Lambda,b) = e^{i \Lambda p \cdot \epsilon(\Lambda,x,b)} \sqrt{\frac{(\Lambda p)^t}{p^t}}  \sum_{\bar{\sigma}} D^{(j)}_{\sigma \bar{\sigma}}(W^{-1})  a_{\bar{\sigma}}({\overrightarrow{\Lambda p}}) \, ,$$
%\end{equation}
\begin{equation} \label{Da+}
U(\Lambda,b) a^{\dagger}_{\sigma}({\overrightarrow{p}}) {U}^{-1}(\Lambda,b) = e^{-i \Lambda p \cdot \epsilon(\Lambda,x,b)} \sqrt{\frac{(\Lambda p)^t}{p^t}}  \sum_{\bar{\sigma}} D^{(j)\ast}_{\sigma \bar{\sigma}}(W^{-1})  a^{\dagger}_{\bar{\sigma}}({\overrightarrow{\Lambda p}}) \, ,
\end{equation}
where $j$ is the spin of the particle and the transformation $W$ is the Wigner rotation given by
\begin{equation} \label{WLp}
W(\Lambda, p) = L^{-1}(\Lambda p)\Lambda L(p) \, ,
\end{equation}
with $L(p) $ a standard transformation taking $k^{\mu}$ = $\{0,0,0,m\}$ to $p.$ $L(p) $ rotates $\overrightarrow{p}$ to the $z$ direction, followed by a boost along $z$, followed by the rotation back to $\overrightarrow{p}.$ 

The unitary transformation $U(\Lambda,b)$ is required to have the effect of a nonunitary transformation on the fields. One requires that 
\begin{equation} \label{Dpsi}
U(\Lambda,b) \psi^{\pm}_{l}(x) {U}^{-1}(\Lambda,b) = \sum_{\bar{l}} D^{-1}_{l \bar{l}}(\Lambda,b)  \psi^{\pm}_{\bar{l}}(\Lambda x + b) \, ,
\end{equation}
where $\Lambda x + b$ are the transformed coordinates of the given event and $D(\Lambda,b)$ is the nonunitary matrix representing the spacetime transformation $(\Lambda,b).$ The rep $D(\Lambda,b)$ is determined by the spin composition of $\psi.$ Let the spin composition of the field $\psi$ be  $(A,B)\oplus(C,D)\oplus \ldots \,.$ Standard angular momentum generators for the nonunitary rep $D(\Lambda,b)$ can be found, for example, in Ref. \cite{rs0}.

We proceed now to the derivation of the coefficients $u$ and $v,$ adapted from Ref. [1].  The coefficients $u$ and $v$ and the operator displacement function $\epsilon(\Lambda,x,b)$ are determined by assuming ({\it{i}}) the operators transform by  $(\Lambda,\epsilon)$ as in (\ref{Da+}) , ({\it{ii}}) the coefficients are invariant,  and ({\it{iii}}) the fields transform by $(\Lambda,b)$  as in (\ref{Dpsi}).  Since the operators $a$ and $a^{\dagger}$ are linearly independent, by (\ref{psi+}), (\ref{Da+}) and (\ref{Dpsi}), the coefficients must obey
   $$ e^{i \Lambda p \cdot \epsilon(\Lambda,x,b)} \sum_{\bar{l}} D_{l \bar{l}}(\Lambda,b) u_{\bar{l}\sigma}(x,{\overrightarrow{p}}) = \sqrt{\frac{ (\Lambda p)^t}{p^t}} \sum_{\bar{\sigma}} u_{l\bar{\sigma}}(\Lambda x + b,{\overrightarrow{\Lambda p}}) D^{(j)}_{\bar{\sigma} \sigma}(W(\Lambda,p))    \, $$
\begin{equation} \label{Du1}
   e^{-i \Lambda p \cdot \epsilon(\Lambda,x,b)} \sum_{\bar{l}} D_{l \bar{l}}(\Lambda,b) v_{\bar{l}\sigma}(x,{\overrightarrow{p}}) = \sqrt{\frac{ (\Lambda p)^t}{p^t}} \sum_{\bar{\sigma}} v_{l\bar{\sigma}}(\Lambda x + b,{\overrightarrow{\Lambda p}}) {D^{(j)}}^{\ast}_{\bar{\sigma} \sigma}(W(\Lambda,p))    \, .
\end{equation}
To solve these equations for the coefficients we consider special cases.

%\pagebreak
As a first case consider translations. Eq. (\ref{Du1}) involves the $u$s and $v$s evaluated at two different sets of coordinates, $x$ and $\Lambda x+b,$ of the given event in two different reference frames. We translate to third and fourth reference frames so that the $u$s and $v$s are evaluated at the same coordinates for the given event. In this case we translate but do not rotate, so the third and fourth reference frames are just as misaligned as the first and second.

For pure translations, $\Lambda$ = 1.  By (\ref{WLp}) $W$ = $L^{-1}(p) L(p)$ = 1. Replacing  $x \rightarrow y,$ $b \rightarrow \bar{b},$ $p \rightarrow q$ in (\ref{Du1}), we have
$$   u_{l\sigma}( y ,{\overrightarrow{ q}}) = e^{-i q \cdot \epsilon(1,y,\bar{b})} \sum_{\bar{l}} D_{l \bar{l}}(1,-\bar{b}) u_{\bar{l}\sigma}(y+\bar{b},{\overrightarrow{q}})    \, $$
\begin{equation} \label{Du2}
   v_{l\sigma}( y ,{\overrightarrow{ q}}) = e^{i q \cdot \epsilon(1,y,\bar{b})} \sum_{\bar{l}} D_{l \bar{l}}(1,-\bar{b}) v_{\bar{l}\sigma}(y+\bar{b},{\overrightarrow{q}})    \, . 
\end{equation}

For simplicity, we assume null coordinates, called the `origin', for the common coordinates of the given event in the third and fourth frames. Therefore, consider (\ref{Du2}) for the subcases (a) $y$ = $x,$  $\bar{b}$ = $-x,$ $q$ = $p$ and (b) $y$ = $ \Lambda x + b,$ $\bar{b}$ = $-(\Lambda x + b),$ $q$ = $\Lambda p.$ One finds by using (a) and (b) that the $u$s and $v$s at $x$ and $\Lambda x +b$ depend on the $u$s and $v$s at the origin and we have, by   (\ref{Du1}),
 $$   e^{+i \Lambda p \cdot [\epsilon(\Lambda,x,b)-\Lambda \epsilon(1,x,-x) +  \epsilon(1, \Lambda x + b,-\Lambda x - b)] }\sum_{\bar{l}} D_{l \bar{l}}(\Lambda,0) u_{\bar{l}\sigma}(0,{\overrightarrow{p}}) = \hspace{5cm}$$ $$\hspace{5cm} \sqrt{\frac{ (\Lambda p)^t}{p^t}} \sum_{\bar{\sigma}} u_{l\bar{\sigma}}(0,{\overrightarrow{\Lambda p}}) D^{(j)}_{\bar{\sigma} \sigma}(W(\Lambda,p))    \, $$
$$e^{-i \Lambda p \cdot [\epsilon(\Lambda,x,b)-\Lambda \epsilon(1,x,-x) +  \epsilon(1,\Lambda x + b,-\Lambda x - b)]}\sum_{\bar{l}} D_{l \bar{l}}(\Lambda,0) v_{\bar{l}\sigma}(0,{\overrightarrow{p}}) = \hspace{5cm}$$ 
\begin{equation} \label{Du4}
     \hspace{5cm} \sqrt{\frac{ (\Lambda p)^t}{p^t}} \sum_{\bar{\sigma}} v_{l\bar{\sigma}}(0,{\overrightarrow{\Lambda p}}) {D^{(j)}}^{\ast}_{\bar{\sigma} \sigma}(W(\Lambda,p))   
\end{equation}
Thus we have coefficients $u$s and $v$s evaluated in reference frames with common, null coordinates for the given event. 

Note that the right-hand-sides of Eq. (\ref{Du4}) do not depend on the coordinates $x$ of the event in the initial reference frame. Nor do they depend on the displacement $b,$ so the left-hand-sides cannot depend on $x$ or $b$ either. 

Since $x$ and $b$ occur on the left only in the operator displacement function $\epsilon,$ one seeks a suitable function $\epsilon(\Lambda,x,b)$ such that the expression in brackets in the phase doesn't depend on $x$ or $b.$ One can show that $\epsilon(\Lambda,x,b)$ must be in the following form,
\begin{equation} \label{epsilon1}
 \epsilon^{\mu}(\Lambda,x,b) =  \epsilon^{\mu}(\Lambda) - \Lambda^{\mu}_{\sigma} M^{\sigma}_{\nu}(x) x^{\nu} + M^{\mu}_{\nu}(\Lambda x+b) (\Lambda x+b)^{\nu} \, ,
\end{equation}
where $M^{\sigma}_{\nu}(x)$ is an arbitrary second rank tensor field defined over the spacetime of allowed coordinates $x$ for the given event. One recovers $\epsilon$ = $b$ when $\epsilon^{\mu}(\Lambda)$ = 0 and the field $M$ is the identity, $M^{\mu}_{\nu}$ = $\delta^{\mu}_{\nu}.$

Assume for convenience, that the four-vector function $\epsilon(\Lambda)$ in (\ref{epsilon1}) is simply $\Lambda \epsilon_{0},$  $$ \epsilon^{\mu}(\Lambda) = \Lambda^{\mu}_{\nu} \epsilon_{0}^{\nu} \, ,$$ for some fixed four-vector $\epsilon_{0},$ independent of $\Lambda,$ $x,$ and $b.$

With the displacement function $\epsilon(\Lambda,x,b)$ in (\ref{epsilon1}), one finds that the left-hand-side of (\ref{Du4}) is $x$- and $b$-independent. One can show that 
\begin{equation} \label{phase}
 \Lambda p \cdot [\epsilon(\Lambda,x,b)-\Lambda \epsilon(1,x,-x) +  \epsilon(1, \Lambda x + b,-\Lambda x - b)] = \Lambda p \cdot \epsilon_{0} \, ,
\end{equation}
which is independent of $x$ and $b$ because the four-vector $\epsilon_{0}$ is independent of $x$ and $b.$

In deriving (\ref{phase}) and in what follows, we need $\epsilon(\Lambda,x,b)$ for $\Lambda$ = 1 and $b$ = $-x;$ i.e. we need $\epsilon(1,x,-x).$ Then $\Lambda x+b$ = 0 and by (\ref{epsilon1}) we have 
\begin{equation} \label{M1a}
 \epsilon^{\mu}(1,x,-x) = \epsilon_{0}^{\mu} - M^{\mu}_{\nu} x^{\nu} \, ,
\end{equation}
where it is understood that the field $M$ is evaluated at $x.$

Continue with this case, i.e. $\Lambda$ = 1 and $W$ = 1, and now with (\ref{M1a}). By (\ref{Du2}) with $q$ = $p,$ $y$ = $x,$ and with $\bar{b}$ = $-x,$ one finds that
   $$ u_{l \sigma}( x,{\overrightarrow{p}}) = 
e^{i p \cdot ( Mx - \epsilon_{0} )} \sum_{\bar{l}} D_{l \bar{l}}(1,x) u_{\bar{l}\sigma}(0,{\overrightarrow{p}})   \, $$
\begin{equation} \label{Du1a}
   v_{l \sigma}( x,{\overrightarrow{p}}) = 
e^{-i p \cdot ( Mx - \epsilon_{0} )} \sum_{\bar{l}} D_{l \bar{l}}(1,x) v_{\bar{l}\sigma}(0,{\overrightarrow{p}})  \, .
\end{equation}
These equations show how the coefficients, the $u$s and $v$s, evaluated at $x$ depend on the $u$s and $v$s evaluated at the origin.

%\pagebreak 

A second case is set up to solve (\ref{Du4}). Let $\Lambda$ = $R$ be a rotation and  $p$ = $k$ = $\{0,0,0,m\}.$ Hence $\Lambda p$ = $Rk$ = $k.$ Since the standard transformation $L(p)$ here transforms $k$ to $p$ = $k,$ one has $L(p)$ = $L^{-1}(Rp)$ = 1 and, by (\ref{WLp}), $W$ = $R.$ 

With the displacement function $\epsilon(\Lambda,x,b)$ in (\ref{epsilon1}), the phase in (\ref{Du4}) reduces to (\ref{phase}) and 
 Eq. (\ref{Du4}) reduces for this case to 
 $$   \sum_{\bar{l}} D_{l \bar{l}}(R,0) u_{\bar{l}\sigma}(0,{\overrightarrow{0}}) = \sum_{\bar{\sigma}} u_{l\bar{\sigma}}(0,{\overrightarrow{0}}) D^{(j)}_{\bar{\sigma} \sigma}(R) \,  $$
\begin{equation} \label{Du5}
     \sum_{\bar{l}} D_{l \bar{l}}(R,0) v_{\bar{l}\sigma}(0,{\overrightarrow{0}}) = \sum_{\bar{\sigma}} v_{l\bar{\sigma}}(0,{\overrightarrow{0}}) {D^{(j)}}^{\ast}_{\bar{\sigma} \sigma}(R) \, ,
\end{equation}
where we assume that $\epsilon_{0}^{t}$ = 0 to make the phase $\Lambda p \cdot \epsilon_{0}$ = $k \cdot \epsilon_{0}$ vanish.
By Schur's Lemma,\cite{H1} either the reps $D$ or $D^{(j)}$ of the group of rotations $R$ are compatible or the coefficients vanish. 

Let the spin composition of the representation matrices $D$ be $(A,B)\oplus (C,D) \ldots .$ Then, by (\ref{Du5}), the coefficients $u(0,{\overrightarrow{0}})$ and $v(0,{\overrightarrow{0}})$ are Clebsch-Gordan coefficients$,^{\cite{W}}$
  $$  u_{l \sigma}(0,{\overrightarrow{0}}) =  \frac{(2 \pi)^{-3/2}}{\sqrt{2m}} \pmatrix{  \langle AaBb \mid j \sigma \rangle \cr  \langle C c D d \mid j \sigma \rangle \cr \vdots   }     \, $$
 \begin{equation} \label{Du6}
    v_{l \sigma}(0,{\overrightarrow{0}}) =  \frac{(-1)^{j+\sigma}(2 \pi)^{-3/2}}{\sqrt{2m}} \pmatrix{  \langle AaBb \mid j, -\sigma \rangle \cr  \langle C c D d \mid j, -\sigma \rangle \cr \vdots   }     \, ,
\end{equation}
where $\langle AaBb \mid j \sigma \rangle$ is the Clebsch-Gordan coefficient which vanishes unless spin $j \in$ $\{\mid A-B\mid, \ldots, A+B\}$ and $a+b$ = $\sigma.$ The factor $1/{\sqrt{2m}}$ is conventional and the factor $(2 \pi)^{-3/2}$ matches the $u$s and $v$s here to those in Ref. \cite{W}. The index $l$ on the left stands for the sequence of double indices $ab,cd, \ldots$ on the right.

A third case is set up to relate the coefficients $u$s and $v$s for momentum $q^{\mu}$ to the $u$s and $v$s for a particle at rest $p^{\mu}$ = $\{0,0,0,m\}.$ 

Let  $\Lambda$ be the special transformation $L(q)$ taking $k$ to $q$ and let $p$ = $k$ = $\{0,0,0,m\}.$ Then, by (\ref{WLp}), $W$ = $L^{-1}(q) L(q) L(k)$ = 1 and, by (\ref{Du4}), one finds that
$$   u_{l{\sigma}}(0,{\overrightarrow{q}}) = e^{i q \cdot \epsilon_{0}}\sqrt{\frac{ m}{q^{\, t}}} \sum_{\bar{l}} D_{l \bar{l}}(L,0) u_{\bar{l}\sigma}(0,{\overrightarrow{0}})     \,$$ 
\begin{equation} \label{Du7}
   v_{l{\sigma}}(0,{\overrightarrow{q}}) = e^{-i q \cdot \epsilon_{0}}\sqrt{\frac{ m}{q^{\, t}}} \sum_{\bar{l}} D_{l \bar{l}}(L,0) v_{\bar{l}\sigma}(0,{\overrightarrow{0}})     \, ,
\end{equation}
where $L$ = $L(q).$ 

By (\ref{Du1a}), (\ref{Du6}) and (\ref{Du7}), one finds an expression for the coefficients $u$ and $v,$
 %\pagebreak
$$   u_{l\sigma}(x ,{\overrightarrow{p}}) = \sqrt{\frac{ m}{p^{\, t}}} \, e^{i p \cdot Mx} \sum_{\bar{l}} D_{l \bar{l}}(L,x) u_{\bar{l}\sigma}(0,{\overrightarrow{0}})     \, ,$$
\begin{equation} \label{Du8}
    v_{l\sigma}(x ,{\overrightarrow{p}}) =  (-1)^{j+\sigma} \sqrt{\frac{ m}{p^{\, t}}} \, e^{-i p \cdot  Mx} \sum_{\bar{l}} D_{l \bar{l}}(L,x) v_{\bar{l}\sigma}(0,{\overrightarrow{0}})     \, ,
\end{equation}
where $L$ = $L(p)$ is the special transformation taking $k$ = $\{0,0,0,m\}$ to $p.$ When $\epsilon_{0}^{t}$ = 0, the  $u(0,{\overrightarrow{0}})$ and the $v(0,{\overrightarrow{0}})$ are the columns of Clebsch-Gordan coefficients in (\ref{Du6}).

The expressions (\ref{Du8}) for the coefficients $u$ and $v$ along with expressions (\ref{psi+}) for $\psi^{+}$ and $\psi^{-}$ determine the quantum field $\psi_{l}(x)$ = $\kappa \psi^{+}_{l}(x)$ + $\lambda \psi^{-}_{l}(x).$ 

For the fields found here, the displacement $\epsilon$ for the annihilation and creation operators differs from the displacement $b$ for spacetime, unless the constant vanishes $\epsilon^{\mu}_{0}$ = 0 and the tensor field $M$ in (\ref{M1a}) is the identity, $M^{\mu}_{\nu}$ = $\delta^{\mu}_{\nu}.$ In that special case, the fields $\psi_{l}(x),$ $\psi^{+}_{l}(x)$ and $\psi^{-}_{l}(x)$ reduce to the conventional free fields that can be found in the literature.\cite{W,S1} 

In this section the spacetime symmetry transformations of the various quantities occurs at a single given event and the value of a quantum field at the given event is found. The next section shows how the active translations considered in Sec. \ref{Phi}, produce quantum fields that respond to forces.

\pagebreak

\section{Dynamics and Classical Limit } \label{ClassLimit}

In this article, momentum changes with translation as well as with rotations and boosts and thereby realizes a nonunitary rep of the Poincar\'{e} group. Forces are introduced implicitly; they are introduced with the nontrivial translation rep described in Sec. \ref{Phi}. As discussed in Sec. \ref{Phi}, an object $\Phi$ with as many as 20 components, including the four components of the momentum, is needed to translate momentum nontrivially. The forces are concealed in $\Phi.$ In this section one sees that the requirement of constant mass leads to an interpretation of the implicit forces as electromagnetism and gravity. 

By introducing forces in the translation rep, the dynamical postulate can be stated as if for free fields.
 
{\it{Dynamical Postulate: A particle in a given eigenstate remains in equivalent eigenstates as spacetime is translated.}}

The idea that spacetime is translated replaces the classical idea that a particle follows a path. In quantum mechanics a particle does not follow one path and in some formalisms all paths contribute. Localizing the particle to a particular path violates the Heisenberg uncertainty principle. But translations displace the entire spacetime, every event is translated to some new event, so there is no violation of the uncertainty principle. 

Yet in order to use the familiar language of classical physics, we note that a confined particle may be described by allowing it a suitably wide range of momenta. Then the translations of all spacetime can be localized to translations of the region of spacetime containing the particle. So a classical path  for our purposes has some thickness, a sequence of translations applied to all of the relevant portion of spacetime inside a `tubular' path.

By (\ref{Du8}), the quantum fields here are sums of `plane waves' in the form $\exp{(\pm i p \cdot M x)},$ 
where $M,$ which is introduced in (\ref{M1a}), is a second rank tensor field. We are thereby lead to assume that the amplitude is proportional to $\exp{( \pm i p \cdot M \delta  x)}$ for a particle in an eigenstate with momentum $p$ to translate through a displacement $\delta x.$  To be well-defined, $\delta x$ must be so small that neither $p$ nor $M$ change significantly along $\delta x.$ So-called polarization effects are ignored. 

Let $\exp{(\pm i \Theta)}$ be proportional to the amplitude for the particle in a particular momentum eigenstate to follow a given sequence of sufficiently short translations that bring an event 1 to an event 2 along a path $C.$ As discussed above, it is the sequence of translations $\delta x_{a}, \delta x_{b}, \ldots$ that is important, events 1 and 2 must be somewhere in the relevant region of spacetime containing the particle. It is assumed that $M$ changes slowly over that region so that the same $\Theta$ occurs with any of the allowed choices for events 1 and 2.

Then one knows from quantum mechanics that $\Theta$ is the change of phase along $C$ from event 1 to event 2,  
\begin{equation} \label{I1}
     \Theta =  \int_{1}^{2} p \cdot M dx = \int_{1}^{2}\eta_{\alpha \beta} p^{\alpha} M^{\beta}_{\sigma} dx^{\sigma}   \, ,
\end{equation}
where $\eta$ is the (flat) spacetime metric, $\eta$ = diag$\{1,1,1,-1\}.$  

Now we find paths with extreme phase. Consider the phase shift $\delta \Theta$ = $p \cdot M \delta x$ over a short displacement $\delta x.$ For convenience, define the four-vector quantity $\delta y \equiv$ $M \delta x$ so that $\delta \Theta$ = $p \cdot \delta y.$ 

Now arrange the reference frame so that $\overrightarrow{p}$ is in the $x^{3}$ direction. Without further loss of generality, let $p$ and $\delta y$ be written as follows
 \begin{eqnarray} \label{I2}
     p^{\alpha} = m\{0,0,\sinh{\chi},\cosh{\chi}\} \hspace{5cm} \cr \cr{\delta y}^{\beta} =  \delta \tau \{\sinh{\mu} \sin{\theta}\cos{\phi}, \sinh{\mu}\sin{\theta}\sin{\phi}, \sinh{\mu}\cos{\theta}, \cosh{\mu} \}  \, ,
\end{eqnarray}
where $\delta \tau \equiv$ $(-\delta y \cdot \delta y)^{-1/2} > 0$ and the parameters $\chi,\mu,\theta,\phi$ are real valued. The quantity $\tau$ has the same units as $\delta x$ and $\delta y$ with $M$ unitless.

With these parameters, the change of phase $\delta \Theta$ along $\delta y$ is 
%\begin{equation} \label{pMdx1} 
$$\delta \Theta = p \cdot \delta y = \eta_{\alpha\beta} p^{\alpha} \delta y^{\beta} = m \delta \tau [\sinh{\chi}\sinh{\mu}\cos{\theta} - \cosh{\chi}\cosh{\mu}] \, . $$
% \end{equation}
The partial derivative of $\delta \Theta$ with respect to $\theta$ is equal to zero when $\theta$ = 0 or $\pi.$ Then the partial of $\delta \Theta$ with respect to $\mu$ vanishes when $\mu$ = $\chi$ or $-\chi,$ respectively. By (\ref{I2}), both choices give the same $\delta y.$ The result is that the extreme phase shift, $\delta \Theta$ = $-m \delta \tau,$ occurs when $\delta y$ is proportional to $p,$
\begin{equation} \label{pMdx2} \delta y^{\alpha} = \delta Y^{\alpha} = M^{\alpha}_{\mu} {\delta X}^{\mu} = m^{-1} p^{\alpha} \delta \tau   \, .
 \end{equation}
Let upper case letters $\delta Y$ and $\delta X$ stand for displacements that make $\delta \Theta$ extreme.

A sequence of extreme phase displacements, e.g. $\{ {\delta X}_{a},{\delta X}_{b}, \ldots \},$ of the relevant region of spacetime produces a `thick' classical path. The collection of such paths determines a coordinate system. We have thus far two such systems of  `extreme coordinates' $X$ and $Y.$ 

By (\ref{pMdx2}) and assuming $M$ is invertible, the tensors $M$ and $M^{-1}$ transform the extreme coordinate systems $ X$ and $Y$ one into the other. We have
\begin{equation} \label{M2}   M^{\alpha}_{\mu} = \frac{\partial{Y^{\alpha}}}{\partial{ X^{\mu}}} \quad {\mathrm{and}} \quad  {(M^{-1})}^{\mu}_{\alpha} = \frac{\partial{X^{\mu}}}{\partial{Y^{\alpha} }} \, .
 \end{equation}
By definition $\delta y$ = $M \delta x,$ so $M$ is also $\partial{y}/\partial{x}.$

 The Wigner class of momenta for a massive particle has $p \cdot p$ = $-m^2$ and $p^{t}\geq m >$ 0, where $m$ is the particle mass. Since $\delta \tau \geq 0,$ Eq.~(\ref{pMdx2}) puts $\delta Y$ in the forward light cone. By varying $\overrightarrow{p},$ the intervals $\delta Y$ fill the forward light cone limited by the value of $\delta \tau.$ Thus classical trajectories for a particle of mass $m$ are confined to the forward light cone because the momentum is restricted by $p \cdot p$ = $-m^2$ and $p^{t}\geq m >$ 0.

By indicating the derivative with respect to $\tau$ with a dot,   i.e. $$\dot{Y} \equiv  dY/d \tau,$$ one can rewrite (\ref{pMdx2}) in a traditional notation,
\begin{equation} \label{Ydot1}  p^{\alpha} = m \dot{ Y}^{\alpha}   \quad {\mathrm{and}} \quad p^{\alpha} = m M^{\alpha}_{\mu} \dot{ X}^{\mu}\, .
 \end{equation}
Thus the `velocity' $\dot{ X}$ is not in the direction of the momentum, while the `velocity' $\dot{ Y}$ is in the direction of the momentum $p.$

Next, incorporate the fact that the (flat) spacetime magnitude of the momentum is the particle mass, $$ \eta_{\alpha \beta} p^{\alpha}p^{\beta} = -m^{2} \, .$$
From this and (\ref{Ydot1}) it follows that the (flat) spacetime magnitude of $\dot{Y}$ is constant,
\begin{equation} \label{g1}  \eta_{\alpha \beta} \dot{ Y}^{\alpha}\dot{ Y}^{\beta} = -1   \, .
\end{equation}
For $\dot{X},$ one finds by (\ref{Ydot1}) that the `curved spacetime magnitude' is constant,
\begin{equation} \label{g2}  g_{\mu \nu} \dot{ X}^{\mu}\dot{ X}^{\nu} = -1    \, ,
\end{equation}
where the `curved spacetime metric' $g_{\mu \nu}$ is defined by
\begin{equation} \label{g3}  g_{\mu \nu} \equiv  \eta_{\alpha \beta} M^{\alpha}_{\mu}M^{\beta}_{\nu} \, .
\end{equation}
Introduced in Eq. (\ref{epsilon1}), $M$ is an arbitrary field, so $g_{\mu \nu}$ is an arbitrary field.

By (\ref{g2}), the quantity $d\tau$ = $\sqrt{-g_{\mu \nu} { dX}^{\mu}{d X}^{\nu}} $ is herein called the `proper time' associated with the curved spacetime metric $g_{\mu\nu}$ along a path of extreme phase.

We note for use below that $g_{\mu \nu}$ has an inverse whenever, as is assumed here, the tensor $M$ has an inverse. We write the inverse of $g_{\mu \nu}$ with raised indices as follows,
%\begin{equation} \label{g3}  
$$g^{\mu \nu} =  \eta^{\alpha \beta} (M^{-1})_{\alpha}^{\mu} (M^{-1})_{\beta}^{\nu} \, .$$
%\end{equation}
Both $g_{\mu \nu}$ and $g^{\mu \nu}$ are symmetric in $\mu$ and $\nu$ because $\eta_{\alpha \beta}$ and $\eta^{\alpha \beta}$ are symmetric in $\alpha$ and $\beta.$

By Assumption 1 and the Dynamical Postulate, the translation of momentum rule (\ref{Dp2}) applies along a path of extreme phase, implying
$\dot{ p}^{\alpha}$ =  $\eta_{\sigma \mu} \bar{T}^{\alpha \sigma }_{(p)}  {\dot{X}}^{\mu}.$ Note the scalar product of $\bar{T}_{(p)}$ and $\dot{X}$ using the (flat) spacetime metric $\eta.$ By defining $T_{(p)}$ (no bar) in
%\begin{equation} \label{Tp1}
 $$  T^{\alpha \mu }_{(p)} \equiv g^{\mu \nu}\eta_{\sigma \nu} \bar{T}^{\alpha \sigma }_{(p)}     \, ,$$
%\end{equation}
one can rewrite equation (\ref{Dp2}) with the curved spacetime metric as follows,
\begin{equation} \label{Dp3a}
\dot{ p}^{\alpha} =  g_{\sigma \mu} T^{\alpha \sigma }_{(p)}  {\dot{X}}^{\mu}  \, .
\end{equation}
This equation for the change in momentum can be turned into an equation for the `four-acceleration' $\ddot{X}$ by defining  the quantity $T^{\mu \sigma }_{(x)}$ as
%\begin{equation} \label{Tx1}
 $$ T^{\mu \sigma }_{(x)} \equiv m^{-1} (M^{-1})_{\alpha}^{\mu} \, T^{\alpha \sigma }_{(p)} - g^{ \sigma \nu} (M^{-1})_{\alpha}^{\mu}\frac{dM^{\alpha}_{\nu}}{d\tau}    \, .$$
%\end{equation}
Then, by substituting the expression (\ref{Ydot1}) for $p$ into (\ref{Dp3a}), one finds that
\begin{equation} \label{D2x1}
\ddot{ X}^{\mu} =  g_{\sigma \nu} T^{\mu \sigma }_{(x)}  {\dot{X}}^{\nu}  \, .
\end{equation}

Since $\bar{T}_{(p)}$ is not constrained when introduced in Sec. \ref{Phi}, $T_{(x)}$ is initially arbitrary. However, by (\ref{Ydot1}) along paths of extreme phase, the relation $p^2$ = $-m^2$ constrains $\dot{X}$ which constrains $\ddot{X}.$ Thus $T_{(x)}$ is not arbitrary along paths of extreme phase. 

The curved spacetime magnitude of $\dot{X}$ is constant along the path of extreme phase by (\ref{g2}). If the derivative is taken with respect to the proper time $\tau,$ i.e. $d(g_{\mu \nu} \dot{ X}^{\mu}\dot{ X}^{\nu})/d\tau$ = 0, and one substitutes (\ref{D2x1}) for the second derivatives of $X,$ one has from (\ref{g2}) that
\begin{equation} \label{DgXX}  \left( \dot{g}_{\mu \nu} + g_{\rho \nu} g_{\sigma \mu} T^{\rho \sigma}_{(x)} + g_{\mu \rho } g_{\sigma \nu} T^{\rho \sigma}_{(x)}\right) \dot{ X}^{\mu}\dot{ X}^{\nu} = 0    \, .
\end{equation}
Without loss of generality, one can introduce quantities $F$ and $\Gamma$ so that $T_{(x)}$ takes the form
\begin{equation} \label{Tx2}  T^{\rho \sigma}_{(x)} = \frac{e}{m} F^{\rho \sigma} - g^{\sigma \kappa} \Gamma^{\rho}_{\kappa \lambda} \dot{X}^{\lambda}    \, ,
\end{equation}
with $F$ made to cancel itself out of (\ref{DgXX}) by requiring that
%\begin{equation} \label{antiF}  
$$F^{\sigma \rho } = - F^{\rho \sigma}    \, .$$
%\end{equation}
By going to a frame in which $\dot{X}$ has only its time component nonzero,  $\dot{X}$ = $\{0,0,0,\dot{X}^{t}\},$ one can see that (\ref{Tx2}) does not constrain $T_{(x)}$ because there are sixteen components of $T^{\mu \nu}_{(x)}$ while there are six free components of the antisymmetric $F^{\sigma \rho }$ and sixteen free components of $\Gamma^{\rho}_{\kappa t}.$

By considering nearby paths of extreme phase one can write the derivative of the metric $g$ with respect to proper time $\tau$ in terms of the $X$ velocity components as follows
\begin{equation} \label{dg}  \dot{g}_{\mu \nu} = \frac{\partial{g_{\mu \nu}}}{\partial{X^{\lambda}}}\dot{X}^{\lambda}    \, .
\end{equation}
Substituting (\ref{Tx2}) and (\ref{dg}) in the expression (\ref{DgXX}) for $d{(g_{\mu \nu} \dot{ X}^{\mu}\dot{ X}^{\nu})}/d\tau$ yields
\begin{equation} \label{DgXX1}  \left( \frac{\partial{g_{\mu \nu}}}{\partial{X^{\lambda}}} - g_{\rho \nu} \Gamma^{\rho}_{(\mu \lambda)}  - g_{\rho \mu} \Gamma^{\rho}_{(\nu \lambda)} \right) \dot{X}^{\lambda} \dot{ X}^{\mu}\dot{ X}^{\nu} = 0    \, ,
\end{equation}
where, by the symmetry evident in the product of the $\dot{X}$s, only the symmetric part of $\Gamma$ contributes. The symmetric and antisymmetric parts of $\Gamma$ are defined to be
%\begin{equation} \label{GammaSymm} 
 $$\Gamma^{\rho}_{(\mu \lambda)} \equiv  \frac{1}{2}\left(\Gamma^{\rho}_{\mu \lambda} + \Gamma^{\rho}_{\lambda \mu }\right)    \quad {\mathrm{and}} \quad  \Gamma^{\rho}_{[\mu \lambda]} \equiv  \frac{1}{2}\left(\Gamma^{\rho}_{\mu \lambda} - \Gamma^{\rho}_{\lambda \mu }\right) 
   \, .$$
%\end{equation}
Since the paths of extreme phase have sufficiently arbitrary tangents $\dot{X}$, the expression in parentheses in (\ref{DgXX1}) vanishes. 

The vanishing of the expression in parentheses in (\ref{DgXX1}) leads by permuting indices $\lambda, \mu, \nu$ to an expression for the symmetric part of $\Gamma.$ One deduces that
\begin{equation} \label{Gamma1}  \Gamma^{\rho}_{(\mu \nu)} =  \frac{g^{\rho \lambda}}{2} \left( \frac{\partial{g_{\lambda \mu}}}{\partial{X^{\nu}}} +\frac{\partial{g_{\nu \lambda}}}{\partial{X^{\mu}}} - \frac{\partial{g_{\mu \nu}}}{\partial{X^{\lambda}}} \right)
   \, .
\end{equation}
Thus the symmetric part of the quantity $\Gamma$ is the Christoffel connection of the curved metric $g.$\cite{Christoffel}

By substituting the expression (\ref{Tx2}) for $T_{(x)}$  in the equation (\ref{D2x1}) for $\ddot{X}$ , one finds that
\begin{equation} \label{D2x2}
\ddot{ X}^{\mu} = \frac{e}{m} g_{\sigma \nu}  F^{\mu \sigma}{\dot{X}}^{\nu} -  \Gamma^{\mu}_{ (\lambda \nu)} \dot{X}^{\lambda} {\dot{X}}^{\nu}  \, .
\end{equation}
This is the equation for the classical trajectory of a particle of mass $m$ and charge $e$ in an electromagnetic field $F^{\rho \sigma}$ and the gravitational field due to a metric $g_{\mu \nu}.$\cite{Somebody}

It should be emphasized that the problem of determining the electromagnetic field $F^{\rho \sigma}$ and the  metric $g_{\mu \nu}$ from the motion of source charges and masses is not considered here. Thus the identification of $F^{\rho \sigma}$ and $g_{\mu \nu}$ with the electromagnetic field and the metric, respectively, is based largely on the resemblance of (\ref{Gamma1}) and (\ref{D2x2}) to equations from general relativity.

One defines the `covariant derivative' of $\dot{X}$ to be
\begin{equation} \label{D2x3}
\frac{D\dot{ X}^{\mu}}{d\tau} \equiv  \ddot{ X}^{\mu} +  \Gamma^{\mu}_{ (\lambda \nu)} \dot{X}^{\lambda} {\dot{X}}^{\nu}  \, .
\end{equation}
Then (\ref{D2x2}) can be written in terms of `covariant' quantities. One has
 \begin{equation} \label{D2x4}
\frac{D\dot{ X}^{\mu}}{d\tau} = \frac{e}{m} g_{\sigma \nu}  F^{\mu \sigma}{\dot{X}}^{\nu}   \, .
\end{equation}
This equation is invariant when the extreme coordinates $X$ are replaced by some other set of extreme coordinates $Z,$ which entails replacing $M$ = $\partial{Y}/\partial{X}$ with some other quantity $N$ = $\partial{Y}/\partial{Z},$ see (\ref{M2}). 

Thus general covariance arises from the freedom to choose $M$ which in turn arises from the arbitrariness assumed for the displacement $\epsilon$ when spacetime is displaced by an amount $b.$ Any displacement of (flat) spacetime preserves all coordinate differences.

\vspace{1.0cm}
\appendix

\section{Problems} \label{problems}

\noindent 1. In one part of Eq.~(\ref{Ydot1}), the momentum is the mass times the velocity, $p^{\alpha}$ = $m \dot{Y}^{\alpha}.$ That has a certain appeal. But the same equation gives the momentum in terms of the $X$ `velocity' as $p^{\alpha}$ = $m M^{\alpha}_{\mu} \dot{X}^{\mu}.$ This would look better with the mass $m$ absorbed in the tensor $M,$ $$p^{\alpha} =  M^{\alpha}_{\mu} \dot{X}^{\mu} \, .$$ Absorb $m$ in $M$ and determine how the other equations and quantities change in this article.

\vspace{0.3cm}
\noindent 2. Suppose $M^{\alpha}_{\mu}$ = $\delta^{\alpha}_{\mu},$ i.e. $M$ is the identity. Find $g_{\mu \nu},$ $T_{(x)},$ and  $\ddot{X}.$ [Note that $m$ is not absorbed by $M$ in this problem, though of course it could be.]

\vspace{0.3cm}
\noindent 3. (a) Check that $J^{\mu \nu}_{11}$ and $J^{\mu \nu}_{22}$ in Sec. \ref{Phi} satisfy the Lorentz commutation rules: $i[J^{\mu \nu},J^{\rho \sigma}]$ = $\eta^{\nu \rho} J^{\mu \sigma} -\eta^{\mu \rho} J^{\nu \sigma}-\eta^{\mu \sigma} J^{\rho \nu}+\eta^{\sigma \nu} J^{\rho \mu}.$
(b) Check the Poincar\'{e} commutation rules for the angular momentum matrices (\ref{gen1}) and the momentum matrices (\ref{gen4}).

\vspace{0.3cm}
\noindent 4. Suppose $M$ is replaced by some other tensor field $N.$ Let $Z$ indicate the extreme coordinates for $N.$ Assume both $M$ and $N$ have inverses. Let the coordinates $x$ of (flat) spacetime be unchanged. (a) Find an expression relating the extreme coordinates $Z$ to the extreme coordinates $X.$ (b) Also find expressions for the quantities $g$, $F,$ and $\Gamma$ found with $N$ in terms of the same quantities found with $M.$ (c) What does (\ref{D2x4}) look like based on $N?$

\vspace{0.3cm}
\noindent 5. In the classical limit, there is no problem in assigning a value of $\Phi$ to each event in spacetime, thereby making a field $\Phi(x),$ see Sec. \ref{Phi}. The field $T_{(\Phi)}(x)$ in (\ref{pT1}) brings in the force fields. For simplicity, let the parameters $C_{i}$ in $P_{12}$ be $C_{2}$ = $-C_{3}$ = $a/2$ and $C_{1}$ = $C_{4}$ = 0. By finding $\tilde{F}$ in terms of the background electromagnetic field $F$ and by appropriately defining the covariant derivative including torsion, show that $T_{(\Phi)}$ can be written in the form
$$ a T^{\sigma \alpha}_{(\Phi)} = e \tilde{F}^{\alpha \sigma} + m \eta^{\sigma \nu} \frac{D}{d \tau} \left(\frac{\partial{Y^{\alpha}}}{\partial{X^{\nu}}} \right) \, . $$

\pagebreak


\begin{thebibliography}{9}

\bibitem{W} S. Weinberg, {\it The Quantum Theory of Fields}, Vol. I (Cambridge University Press, Cambridge, 1995), Chapter 5 and references therein.

\bibitem{Feynman} R. P. Feynman and A. R. Hibbs, {\it Quantum Mechanics and Path Integrals} (McGraw-Hill Book Co., New York, 1965).

\bibitem{rs0} R. Shurtleff, on-line article, {\it arXiv:math-ph/0401002}, 2004.

\bibitem{L} The condition also applies to vector matrices $\beta$ that arise in the well-studied first order equation $\beta^{\mu} \partial_{\mu} \psi$ = $m \psi.$ See, for example, G. Ya. Lyubarskii, {\it The Application of Group Theory in Physics}, trans. by S. Dedijer(Pergamon Press, Oxford, 1960), Chapter XVI.

\bibitem{WeinbergV} S. Weinberg, {\it Gravitation and Cosmology} (John Wiley and Sons, Inc, New York, 1972), Chap. 2, Sec. 12.

\bibitem{RS} R. Shurtleff, on-line article, {\it arXiv:gr-qc/0502021}, 2005.

\bibitem{S1} The work of Ref. \cite{W} was adapted slightly for fields with Poincar\'{e} transformations in R. Shurtleff, on-line article, {\it arXiv:hep-th/0401119}, 2004.

\bibitem{H1} See, e.g., Hamermesh, M., Group Theory and its Application to Physical Problems (Addison-Wesley Publishing Co., Reading, Massachusetts, 1964), Sec.~3-14.

\bibitem{Christoffel} R. Adler, M. Bazin and M. Schiffer, {\it Introduction to General Relativity} (McGraw-Hill Book Co., New York, 1965).

\bibitem{Somebody} C. W. Misner, K. S. Thorne and J. A. Wheeler, {\it Gravitation} (W. H. Freeman and Co., San Francisco 1973), Sec. 22.4.


 
\end{thebibliography}
\end{document}